# Towards identification of a non-abelian state: observation of a quarter of electron charge at $\nu$=5/2 quantum Hall state

M. Dolev, M. Heiblum, V. Umansky, Ady Stern, and D. Mahalu
*Braun Center for Submicron Research, Dept of Cond. Matter Physics,
Weizmann Institute of Science, Rehovot* 76100*, Israel*

**The fractional quantum Hall effect, where plateaus in the Hall resistance at values of $h/\nu e^2$ coexist with zeros in the longitudinal resistance, results from electron correlations in two dimensions under a strong magnetic field. Current flows along the edges carried by charged excitations (quasi particles) whose charge is a fraction of the electron charge. While earlier research concentrated on odd denominator fractional values of $\nu$, the observation of the even denominator $\nu$=5/2 state sparked a vast interest. This state is conjectured to be characterized by quasiparticles of charge *e*/4, whose statistics is "non-abelian". In other words, interchanging of two quasi particles may modify the state of the system to an orthogonal one, and does not just add a phase as in for fermions or bosons. As such, these quasiparticles may be useful for the construction of a topological quantum computer. Here we report data of shot noise generated by partitioning edge currents in the $\nu$=5/2 state, consistent with the charge of the quasiparticle being *e/*4, and inconsistent with other potentially possible values, such as *e*/2 and *e*. While not proving the 'non-abelian' nature of the $\nu$=5/2 state, this observation is the first step toward a full understanding of these new fractional charges.**

Theoretical predictions regarding the nature of the even denominator $\nu$=5/2 quantum Hall state rekindled strong interest in the fractional quantum Hall state (FQHE) [1,2,7]. Primarily, this interest emanates from the unique properties for quasiparticles (qp's) in this state, predicted by Moore and Read [8]: a fractional charge of a quarter of the electron charge (*e*/4) and a non-abelian quantum statistics. The quantum statistics is reflected in the evolution of the ground state wave function when two *e*/4 qp's are adiabatically interchanged. For conventional FQHE states, where the statistics is abelian, such an interchange merely multiplies the wavefunction by a phase. For non-abelian states, the presence of qp's makes the ground-state degenerate and an adiabatic interchange of two qp's leads to a topological unitary transformation - where the topology of the path determines the transformation - that takes the system from one ground state to another. Unitary transformations that correspond to different interchanges do not generally commute with each other; hence the name non-abelian.

The topological nature these transformation makes the $\nu$=5/2 state a test ground of the basic ideas of topological quantum computation, as it introduces remarkable immunity against decoherence and errors due to local uncontrollable perturbations [10,11,12]. Ideas proposed in these directions are based interference experiments where inter-edge tunnelling of *e*/4 qp's takes place. For these experiments to succeed the $\nu$=5/2 state must be of the Moore-Read type and the tunnelling qp's must have charge *e*/4. Presently, these characteristics are predicted but not yet experimentally confirmed. The Moore-Read theory is based on a trial wave function inspired by considerations of conformal field theory [8]. It may be rederived [13] by considering weak Cooper pairing of composite fermions, which under a magnetic field corresponding to the $\nu$=5/2 state, are fermions carrying an electron charge and two fictitious flux quanta. It is also supported by numerical exact diagonalization [17]. As for the second requirement, if electron or $e/2$ qp tunnelling dominates over that of the $e/4$ qp's, inter-edge tunnelling would not be a useful tool for examining the non-abelian statistics of the latter.

In this work we present shot noise measurements [5,6], which result from partitioning of a stream of qp's that tunnel between edge channels of a $\nu$=5/2 state. Measurements were performed on a patterned high purity two dimensional electron gas with a built in constriction that allowed controlled tunnelling from one edge to another. From the dependence of the shot noise on the current we deduced the charge of the qp's, which was found to be consistent with charge $e/4$, and inconsistent with charge $e/2$ or an electron charge $e$. In order to further validate the measurements, the charge of qp's in other FQHE states in the vicinity of the $\nu$=5/2 state, such as at $\nu$=5/3, 2, 8/3, & 3, was also measured by partitioning their current carrying states. While these measurements do not directly probe non-abelian statistics, they do pave the way for such a measurement, which is likely to be based on interference effects [26,27,28,29].

**The 5/2 fractional quantum Hall state**

The $\nu$=5/2 state is characterized by a zero longitudinal conductance and a Hall conductance plateau $g_{5/2} = \dfrac{5e^2}{2h}$ ($h$ the Planck constant). At $\nu$=5/2, the highly interacting electronic system can be mapped onto a system of weakly interacting, spin polarized, composite fermions at zero average magnetic field. The theory that predicts the $\nu$=5/2 state to be non-abelian starts with the Moore-Read (MR) wave function [8]. Within this theory the composite fermions form a superconductor of Cooper-pairs, with $p_x+ip_y$ symmetry [13], where the non-abelian qp's are vortices, carrying half of a quantum flux $\phi_0/2 = h/2e$. Since inserting these vortices costs a finite amount of energy (the Meissner effect in the condensate), the system is in an incompressible quantum Hall state with an energy gap for these excitations. With the flux carried by a vortex being $h/2e$ and the uppermost Landau level half filled, the charge that is associated with the qp's was theoretically predicted to be $e^* = e/4$ [8]. However, recent numerical calculations raise the possibility that these qp's may tend to form bound pairs of charge $e/2$ [14]. Only a small number of studies tested thus far the predictions of Moore-Read theory [8]. The observed energy gap was measured to be significantly smaller than predicted [15]; the survival of the state in a small constriction was studied [16]; and spin polarization of the state is thus far supported only by a numerical exact diagonalization method [17]. What makes measurements at the $\nu$=5/2 state rather scarce is the extreme fragility of the state. Only extremely high quality two dimensional electron gas (2DEG) and a rather low electron temperature support such states.

**Setup and the basics of measurements**

The most important ingredient in our measurements is the GaAs-AlGaAs heterostructure that supports the 2DEG. The 2DEG, with an electron density of $\sim 3.2 \times 10^{11}$cm$^{-2}$, was confined in a 29nm wide quantum well, with a low temperature mobility of $30.5 \times 10^6$cm$^2$/V-s, measured in the dark (see Methods for more details). The quantum scattering time of the electrons was estimated from Shubnikov de Haas oscillations to be around 12ps. The Hall effect data, measured at 10mK in a Hall bar 1mm wide, revealed five significant fractions in the second Landau level, $\nu$=11/5, 7/3, 5/2, 8/3, & 14/5, with a nearly zero longitudinal resistance at $\nu$=5/2 (Fig. 1a). This data indicates transport via edge channels with negligible current flowing in the bulk.

The samples were patterned in a shape of a Hall bar, with a single quantum point contact (QPC) in its center and multiple ohmic contacts (see inset in Fig. 1b). The QPC partitioned the incoming current by bringing the forward propagating edge channel, with a chemical potential determined by the applied voltage $V$, to close proximity with the backward propagating edge channel returning from the grounded contact, hence inducing backscattering, partitioning, and shot noise. Two contacts, 1 and 2, were grounded directly to the cold finger ('cold grounds'), which was attached directly to the mixing chamber, in order to cool the electrons to ~10mK [18]. When needed, a heater was used to heat the mixing chamber, and hence the sample, above base temperature. A DC current $I_{imp}$ was driven from the source contact ($S$), with the QPC partitioning it to the transmitted and backscattered currents. Shot noise, at 910kHz, was amplified by a home-made preamplifier cooled to 4.2K, with its output fed into a room temperature amplifier followed by a spectrum analyzer (see Method for more details). The multi terminal configuration of the sample ensured a constant input resistance to the preamplifier when the QHE was tuned to a conductance plateau; independent of the transmission of the QPC (As long as the longitudinal resistance was zero the resistance between terminals D & 4 was always the Hall resistance) [19]. This allowed the 'resistance dependent' noise components, namely, the 'current noise' of the preamplifier (backward injected current by the preamplifier into the sample) and the thermal noise of the sample (together being some 50 times larger then the desired shot noise signal), to be subtracted from the measured total signal (their summed value was measured by setting the impinging current to zero).

We measured the two-terminal conductance and the current noise as a function of the source-drain voltage, at different partitioning values set by the QPC. Performing the noise measurements turned out to be challenging. A severe difficulty during the measurements was the instability and irreproducibility of the gates. Consequently, the desired partitioning by the QPC, which was determined by the filling factor within the QPC, was achieved by controlling the gate voltage (which controls the density within the QPC) simultaneously with tuning the magnetic field along a conductance Hall plateau in the bulk. This difficulty was compounded by the fact that the shot noise signal at the $\nu = 5/2$ state is very weak. Shot noise current fluctuations $S^i$ are proportional to the effective charge $e^*$ and to the impinging current in the partitioned edge channel. At $\nu = 5/2$ the relevant partitioned current, which generates the noise, is only a small fraction of the total impinging current, depending on the next-lower channel within the QPC (e.g., if the lower channel is $\nu$=2, only 1/5 of the impinging current generates the noise). Furthermore, since our "cold preamplifier" amplifies voltage fluctuations $S^V$, and $S^V=S^i/(g_{5/2})^2$, the large conductance at $\nu = 5/2$ leads to a rather weak voltage signal.

**The analysis of shot noise**
For the extraction of the qp's charge we employ a rather simplified picture which was successful in analyzing the shot noise of qp's with charge $e/3$ at $\nu$=1/3, $e/5$ at $\nu$=2/5, $e/7$ at $\nu=3/7$ [5,20,18]. In this picture we assume that in a conductance Hall plateau the current is being carried by qp's in chiral edge channels with no bulk currents. When the bulk is in a certain conductance plateau and the QPC is wide open, the edge currents, which are noiseless (at zero temperature), traverse the QPC unaltered. As the QPC starts to close, the innermost forward propagating edge channel is coupled to the innermost backward propagating edge channel via tunneling, leading to statistically independent backscattering events. Pinching the QPC further leads eventually to a full reflection of this partitioned edge channel with the conductance through the QPC reaching a lower conductance plateau.

When the conductance dependent current through the QPC $g(I_{imp})$ is in a transition between two plateaus, one corresponding to a lower lying state $\nu_{i-1}$ with conductance $g_{i-1} = \nu_{i-1}e^2/h$ and another to the state above $\nu_i$ with conductance $g_i = \nu_i e^2/h$, the current that impinges on the innermost channel is $I_{imp}(i)=V\Delta g_i$ with $\Delta g_i = g_i - g_{i-1}$, where $V$ is the applied DC voltage at the source. This current, which is in general smaller than the total impinging current $I_{imp}$, is being partitioned hence generating shot noise. Similarly, the transmission of this channel's current is defined as $t_{\nu_i-\nu_{i-1}} = \dfrac{g(I_{imp})-g_{i-1}}{\Delta g_i}$. At zero temperature ($T=0$), the low frequency spectral density of the current fluctuations $S^i(0)$ is related to the quasi-particle's effective charge $e^*$ through [21]:

$$S^i(0)_{T=0} = 2e^* V \Delta g_i\, t_{\nu_i-\nu_{i-1}}(1-t_{\nu_i-\nu_{i-1}}) \quad . \tag{1}$$

At finite temperatures the thermal contribution adds in and the shot noise is being modified to the so called 'excess noise' [22]. The expression in Eq. (1) is being then modified [22]:

$$S^i(0)_T = 2e^* V \Delta g_i\, t_{\nu_i-\nu_{i-1}}(1-t_{\nu_i-\nu_{i-1}})\left[\coth\left(\dfrac{e^*V}{2k_BT}\right) - \dfrac{2k_BT}{e^*V}\right] + 4k_BTg \quad , \tag{2}$$

with $k_B$ the Botzmann constant. The assumption of statistical independence of the backscattered qp's is expected to hold when $t_{\nu_i-\nu_{i-1}}$ is close to zero (with rarely transmitted qp's) or close to unity (with rarely reflected qp's), but may be less reliable in intermediate values of $t_{\nu_i-\nu_{i-1}}$. Yet, previous works on Laughlin's qp's have proven that if the non-linear transmission is taken into account, then the prediction of the qp's charge from Eqs. (1) & (2) agrees with the expected one. This might be attributed to the fact that Eq. (2) does have the correct limits at low and high temperatures, as well as the correct scale for the transition between the two.

**Identification of the structure of edge channels in the QPC**
Deducing the charge from the measured shot noise (via Eq. 2) requires a measurement of the relevant transmission $t_{\nu_i-\nu_{i-1}}$ (defined above) and the corresponding impinging current $I_{imp}(i)$, which in turn requires the identification of the next-lower quantized Hall conductance $g_{i-1}$ (the next to innermost edge channel) within the QPC. Our procedure for identifying the occupied channels in the QPC is demonstrated in Fig. 2. At low temperature the transmission of the QPC strongly depends on the impinging DC current. This dependence does not generally agree with the prediction of the chiral Luttinger liquid (CLL) theory. As had been already observed before, in the integer and fractional quantum Hall regime [18,23,24], under weak backscattering conditions of the $i^{th}$ state (the QPC is tuned to just below the $i^{th}$ plateau in the bulk, such that the $i^{th}$ edge channel is weakly backscattered), the transmission decreases with increasing DC current, exhibiting 'mound-like' behavior. Alternatively, for strong backscattering conditions (the QPC is tuned to almost completely pinch off the $i^{th}$ state and the transmission is just above the lower lying $(i-1)^{th}$ plateau, such that the $i^{th}$ channel is almost fully backscattered), the transmission increases with DC current, exhibiting 'valley-like' behavior. We attribute the "mound-like" and 'valley-like' behaviors to a combination of several factors. Generally speaking, a larger applied voltage on the QPC barrier is expected to enhance the bare tunneling probability for independent particles. 'Mound-like' behavior appears in the limit of $t_{\nu_i-\nu_{i-1}}$ close to unity, where tunnelling is between the forward and backward propagating channels. Then, tunneling is responsible for the weak backscattering and thus applying larger voltage across the QPC increases backscattering and decreases the transmission. In contrast, 'valley-like' behavior appears in the limit of $t_{\nu_i-\nu_{i-1}}$ close to zero, where the high barrier is for forward tunnelling over the QPC; hence, tunneling is responsible for the small transmission, which increases with the impinging current. The renormalization of the tunneling rates predicted by the CLL presumably co-exists with this mechanism. However, CLL alone cannot explain this reproducible 'mound-valley' behavior, as it predicts enhancement of the tunneling probability due to a decrease in the applied voltage. Altogether, as the QPC closes and the conductance crosses a quantized plateau, the transmission dependence on the impinging current switches from 'valley-like', to 'flat-like', and finally to 'mound-like' behavior, offering us a way to identify the plateau that is being crossed.

We demonstrate this evolution in Fig. 2a, where we plot the dependence of the transmission on the total impinging DC current $I_{imp}$ at bulk filling factor $\nu_B=3$. The filling factor within the QPC, $\nu_{QPC}$, was varied either by the applied gate voltage $V_g$ or by varying slightly the magnetic field within the bulk plateau of $\nu_B=3$. One can follow the evolution from $\nu_{QPC}=3$ to $\nu_{QPC}=7/3$, with a dependence on the impinging current as alluded above. In Fig. 2b, we plot the dependence of the linear transmission (at zero DC current) on the filling factor in the QPC (the measurement was taken at two different gate voltages with a varying magnetic field at each gate voltage). Clear plateaus were observed within the QPC at $\nu_{QPC}=5/2$ and $\nu_{QPC}=7/3$, and a weaker one at $\nu_{QPC}=8/3$, in agreement with the values in fig. 2a at which a transition was observed. Moreover, at these plateaus no shot noise had been measured at a finite DC current.

The identification of a plateau that corresponds to a certain filling factor in the QPC under certain gate voltage and magnetic field almost guarantees that this filling factor exists, as a lower lying state or as an outer edge channel, at higher values of QPC conductance. This is since at a less negative gate voltage (or a lower magnetic field) the QPC is more open, thus resembling the bulk and allowing the existence of incompressible regions within the QPC. However, the absence of such a plateau near an expected filling factor cannot rule out the existence of this lower lying state when the QPC is more open.

In some cases, at the lowest temperature of 10mK the transmission of the QPC strongly depended on the impinging DC current, making data interpretation difficult. Hence, measurements were also performed at 40mK and at 90mK, where the transmission dependence on the current weakens and the excess noise agrees with Eq. (2)

(see Methods for more details), suggesting an almost single particle-like behavior [see also 5,19,20,23]. The non-linear transmission was taken into account utilizing two different models for the effective transmission $t_{\nu_i-\nu_{i-1}}$: a *differential* model, where for each value of the total impinging current, $I_{imp}$ (we drop the *tot* for simplicity) we take $t_{\nu_i-\nu_{i-1}} \equiv t_{diff}(I_{imp}) = \frac{g(I_{imp}) - g_{i-1}}{\Delta g_i}$; and an *average* model, $t_{\nu_i-\nu_{i-1}} = t_{aver}(I_{imp})$, where the transmission is obtained by an average of the differential transmissions in the current range 0-$I_{imp}$. While in the differential model the underlying assumption is that the potential barrier in the QPC is affected by the applied voltage; in the average model we assume implicitly that the potential barrier in the QPC is independent of the applied voltage. While each model is not accurate, they represent a lower and an upper limit of the transmission of the QPC.

Measurements of shot noise were performed in a wide range of filling factors in the QPC ($\nu_{QPC}$=5/3....3), while in the bulk the filling factor was kept at $\nu_B$=2, $\nu_B$=5/2 or at $\nu_B$=3. The corresponding charge of the qp's in state $\nu_i$ is best measured at the weak backscattering limit of the state $\nu_i$, since one expects then rare and independent backscattering events, and moreover, a weaker dependence of the effective transmission on the correct identification of the next lower lying channel $\nu_{i-1}$. Note that while Eq. (2) should also be valid at small transmission, the charge of the qp's may change in this limit, reflecting already the nature of the next lower lying state $\nu_{i-1}$ [25]. In our setup, a weak and persistent reflection by the QPC, even at an applied zero gate voltage, prevented reaching a sufficiently small backscattering coefficient at $\nu_B$=5/2, hence, we conducted also measurements at bulk filling $\nu_B$=3 with the QPC tuned to filling factors in the range $\nu_{QPC}$=7/3...5/2, as we describe below.

**Shot noise measurements and charge determination**

Conductance and shot noise measurements were conducted at different bulk filling factors based on our identification method of the lower lying states in the QPC described above. As will be shown below, we also tested, when in question, the consequence of a different choice of the next lower lying channel.

We started with measurements of a partitioned 5/2 state within the QPC while the bulk filling factor was $\nu_B$=3. The conductance and shot noise are plotted as function of the total impinging DC current $I_{imp} = 3Ve^2/h$ at a temperature of 40mK (Fig. 3). The impinging current and the differential transmission were calculated under the assumption that the lower channel is $\nu_{QPC}$=7/3 (see measurement results in Fig. 2), namely, $I_{imp}(i) = V\Delta g_{5/2} = V\frac{e^2}{6h}$ and $t_{5/2-7/3} = \frac{g(I_{imp}) - g_{7/3}}{\Delta g_{5/2}}$ =0.55-0.72 (as the impinging current increases). The two models for the effective transmission ($t_{diff}$ & $t_{aver}$) led to similar predictions for the expected noise, with that for charge $e^*$=e/4 plotted in Fig. 3b (black line). While the scattering of the data was relatively large (even after 48 hours of measurement time) the agreement with $e^*$=e/4, excluding thus $e^*$=e/2 (also plotted for comparison), is evident. We note that assuming a next lower lying channel $\nu_{QPC}$=2 does not significantly change our conclusion. In the latter case, the effective transmission varies in the range 0.86-0.90 and the analysis of the shot noise agrees with a qp charge $e^*=e/5$-$e/4$ - definitely excluding a charge $e^*=e/2$.

In order to further study the qp's charge, we performed also measurements at bulk filling factor $v_B=5/2$ and different QPC transmissions. Measurements of the non-linear conductance (similar to Fig. 2) were performed in order to verify the next lower lying channel. They revealed that for these parameters the next lower lying state in the QPC was $v_{QPC}=2$. Moreover, the absence of the $v_{QPC}=7/3$ lower channel was also verified, first, by not observing a flat-like transmission as a function of impinging current, and second, by the absence of noise suppression when the conductance corresponded to $v_{QPC}=7/3$. Figure 4 displays three measurements of conduction and shot noise as a function of the total impinging current $I_{imp}$: Fig. 4a, 4b - with measurements at 10mK and at a reasonably weak backscattering; Fig. 4c, 4d - with measurements at 10mK and transmission $t_{5/2-2} \sim 0.5$; and Fig. 4d, 4e - with measurements at 40mK and very strong backscattering. For weak backscattering (Fig. 4a, 4b), both models for the effective transmission coincide, leading to qp's charge $e^* = e/4$ (the curve for $e^* = e/2$ was also shown for comparison). Figure 4d presents data where the reliability of Eq. (2) is questionable (since the transmission is intermediate, the scattering events may not be independent), and the two models for transmission clearly deviated from each other (shown in the figure). Here, the apparent charge is again close to $e^* = e/4$ although the data fits better to a charge $e^* \approx 0.2e$. When the 5/2 channel is almost completely pinched at zero impinging current (Fig. 4e, 4f), and the transmission is highly non-linear and changes from 2% to 20% as function of the current, again both models for the transmission provide an upper and lower limit for the shot noise, but yet, corresponding nicely to a qp charge $e^* = e/4$. This is somewhat surprising since one might have expected at this strong backscattering regime, with the next lower lying channel $v_{QPC}=2$, that the tunneling particle would be an electron.

It is obviously desirable to measure the effective charge at different quantum Hall states in the same device in the vicinity of $v_{QPC}=5/2$. Figure 5 summarizes some of the measurements we did in the range $v_{QPC}=5/3…..3$, identified on the two terminal Hall conductance of the actual sample where the measurements were conducted on. In the top panel we show the effective transmission, of the state under study with the identification of the next lower lying state, which was identified by the method shown in Fig. 2 (in some of the cases we did not plot the conductance and shot noise due to lack of space). The first panel on the left, measured at $v_B=3$, with relatively weak backscattering induced by the QPC. The transmission drops with increasing current in a 'mound like' fashion and saturates around 5nA. The saturation at higher currents (or voltage) is typical (see, e.g. [23]) and is accompanied by a saturation of the shot noise. It could be related to the deforming shape of the barrier or interactions (charging) induced by the increased current. Here, near in $v_{QPC}=3$, we observed a nice agreement with charge $e$, as expected. Pinching the QPC further, as in the second panel, the conductance approaches the corresponding one to $v_{QPC}=8/3$, the charge fractionalizes with a clear qp charge $e^* = e/3$. A very similar behavior is observed when the filling factor in the QPC is slightly lower than $v_{QPC}=2$ (measured at $v_B=2$ and at a temperature of 90mK), agreeing with a charge $e$, as expected. This charge fractionalized to $e^* = e/3$ as the filling factor lowered to just above $\nu_{QPC}$=5/3. Such fractionalization of the charge as the effective barrier for electrons increases was surprising at first. We attribute it to the formation of a FQHE region of $\nu_{QPC}$=5/3 within the QPC, separating the two regions on both sides of the QPC. Then, the observed fractional charge may be viewed as the charge of the quasi-holes with charge $e^* = e/3$ tunneling across the $\nu_{QPC}$=5/3 region. When the filling factor in the QPC drops further to slightly below $\nu$=5/3, and similarly also below $\nu$=8/3, we also measured $e^* = e/3$ as was expected [5], (see panels in Fig. 5).

**Discussion**
In summary, we presented direct evidence of quasiparticle charge of $e^*$=$e$/4 at an even denominator fractional filling of 5/2 in the second Landau level of the quantum Hall effect. The affirmation of the predicted charge of the quasiparticles is a strong indication that the $\nu$=5/2 is a paired state, where pairs of composite fermions condense into a gapped state. It is consistent with the Moore-Read [8] theory, and indicates that if this theory is indeed correct, the quasiparticles that tunnel across a point contact are the non-abelian quasiparticles with charge $e$/4. Our experiment does not probe the non-abelian statistics. In order to probe this statistics, a direct measurement, say, via interference [26,27,28], should be conducted. Finally, we also measured the charge of quasiparticles in adjacent filling factors.

**METHODS SUMMARY**
**The 2DEG and the sample**
Our structure consisted of 29nm wide GaAs-Al$_{0.25}$Ga$_{0.75}$As quantum well, doped on both sides with 'Si delta-doping'. The two 'delta-doping' were placed in narrow quantum wells, being part of a short period superlattice, separated from the 2DEG by an undoped Al$_{0.25}$Ga$_{0.75}$As with thickness 80nm. The 2DEG was located 160nm below the surface, with the AlAs mol-fraction rising to 35% near the surface.

The sample was patterned by optical lithography and electron beam lithography. Measurements were done in unilluminated sample. The conductance of the QPC was found to be irreproducible as a function of the gate voltage, and tended to vary as function of time after the gate voltage was changed.

Note that the 2DEG mobility and the quantum scattering time were found to be poor indicators for the 'quality of FQHE features'. Some lower mobility samples (around 15×10$^6$cm$^2$/V-s) showed nice fractional states while higher mobility samples (as high as 36×10$^6$cm$^2$/V-s) showed sometimes poorer behavior.

**Measuring shot noise**
By having multi terminal configuration the sample conductance in a QH plateau was constant $\nu e^2/h$ and independent of the QPC transmission. To avoid the large 1/$f$ noise at low frequencies, a resonant circuit was connected between the drain (*D*) and ground, made of a copper coil (*L*) and a capacitor (*C*), tuned to a resonance frequency of 910kHz. This followed by a home-made preamplifier cooled to 4.2K, a room temperature amplifier (NF SA-220F5), and a spectrum analyzer (bandwidth 30kHz or 100kHz).

Comparing the expected spectral density of the voltage noise $S^V$ at $\nu$=1/3 and at $\nu$=5/2, we find $\dfrac{S^V_{1/3}}{S^V_{5/2}} = 50$. Since the signal (the shot noise) to noise (the uncorrelated system noise) ratio is proportional to $(\Delta f \times \tau)^{-1/2}$, with $\Delta f$ the bandwidth, an unreasonable 2500 times longer measurement time $\tau$ is required for the same signal to noise ratio. Hence, a larger voltage was applied (10-100µV, being some 10-100 times larger then the temperature), combined with a wider bandwidth of the $LC$ circuit at the $\nu$=5/2 state (due to the higher conductance) it enabled a more reasonable measurement time.

**Acknowledgments**
We wish to acknowledge A. Ra'anan for laying the grounds for that experiment. We thank for the help and willing assistance provided by A. Schreier, I. Neder, N. Ofek, Y. Gross, E. Grosfeld, Y. Gefen, B. I. Halperin, Y. Levinson and B. Rosenow. We thank J. Miller and C. Marcus for sharing with us their experience on the fabrication process, and to L. Pfeiffer for helping us to start these experiments by providing us with a sample for initial experimentation. MH wishes to acknowledge for partial support the Israeli Science Foundation (ISF), The German Israeli Foundation (GIF), and the MINERVA foundation. AS wishes to acknowledge the US-Israel Bi-national Science Foundation, the Minerva foundation, and the Israel Science Foundation (ISF).

# Methods

**The 2DEG**

The enabling ingredient for the experiment is the quality of the 2DEG. Only extremely high mobility structures showed the fragile even denominator 5/2 fraction. Our heterostructure consisted of 29nm wide GaAs-$Al_{0.25}Ga_{0.75}As$ quantum well containing the 2DEG, doped on both sides with 'Si delta-doping' and serving as donors. The two 'delta-doping' were also placed in narrow quantum wells, being part of a short period superlattice, separated from the 2DEG by an undoped $Al_{0.25}Ga_{0.75}As$ (spacer) with thickness 80nm. Uniform doping, some 60nm below the surface, compensated the surface states and thus terminated the depletion layer. The 2DEG was located 160nm below the surface, with the AlAs mol-fraction rising from 25% at the uniform doping to 35% near the surface. A thin GaAs cap terminated the structure in order to prevent oxidation and facilitate better ohmic contacts.

**The mobility of the 2DEG**

Note that the mobility of 2DEG and the quantum scattering time were found to be rather poor indicators for the 'quality of FQHE features'. Indeed high mobility 2DEG are needed to observe the fragile fractional states, however, in this high range some lower mobility samples (around $15\times10^6 cm^2/V$-s) showed nice fractional states while higher mobility samples (as high as $36\times10^6 cm^2/V$-s) showed sometimes poorer behavior. This behavior is poorly understood, however, since the mobility and quantum times are representing only the second order correlator of the potential fluctuations, it is likely that higher order correlators (namely, the detailed shape of the potential landscape) are responsible for the localization in high magnetic field.

**The sample**

The sample was patterned by optical lithography and electron beam lithography. Metallic gates were formed by deposition of 15nm of PdAu and 15nm of Au. Measurements were done in unilluminated sample. The conductance of the QPC was found to be irreproducible as a function of the gate voltage, and tended to vary as function of time after the gate voltage was changes. Hence, the desired conductance was achieved by tuning the gate voltage and the magnetic field along the QH plateau.

**Measuring shot noise**

By having multi terminal configuration (with grounded contacts on both sides of the QPC) the sample conductance in a QH plateau, viewed by the source or by the drain (connected to the preamplifier) was constant ($ve^2/h$) and independent of the transmission of the QPC. To avoid the large $1/f$ noise at low frequencies, a resonant circuit was connected between the drain (*D*) and ground, made of a cupper coil (*L*) and a capacitor (*C*, formed mostly by the capacitance of the coax cable), tuned to a resonance frequency of 910kHz. This was followed by a home-made preamplifier cooled to 4.2K, a room temperature amplifier (NF SA-220F5), and a spectrum analyzer (bandwidth 30kHz or 100kHz).

Comparing the expected spectral density of the voltage noise $S^V$ at $\nu=1/3$ and at $\nu=5/2$, we find $\dfrac{S^V_{1/3}}{S^V_{5/2}} = 50$. Since the signal to noise ratio (the shot noise to the uncorrelated system noise ratio) is proportional to $(\Delta f \times \tau)^{-1/2}$, with $\Delta f$ the bandwidth and $\tau$ the measurement time, an unreasonable measurement time (2500 times longer) was required for the same signal to noise ratio. Hence, a larger voltage (and current) was applied (10-100µV, being some 10-100 times larger then the electron temperature), which combined with a wider bandwidth of the $LC$ circuit at the $\nu=5/2$ state (due to the relatively high conductance) enabled a more reasonable measurement time.

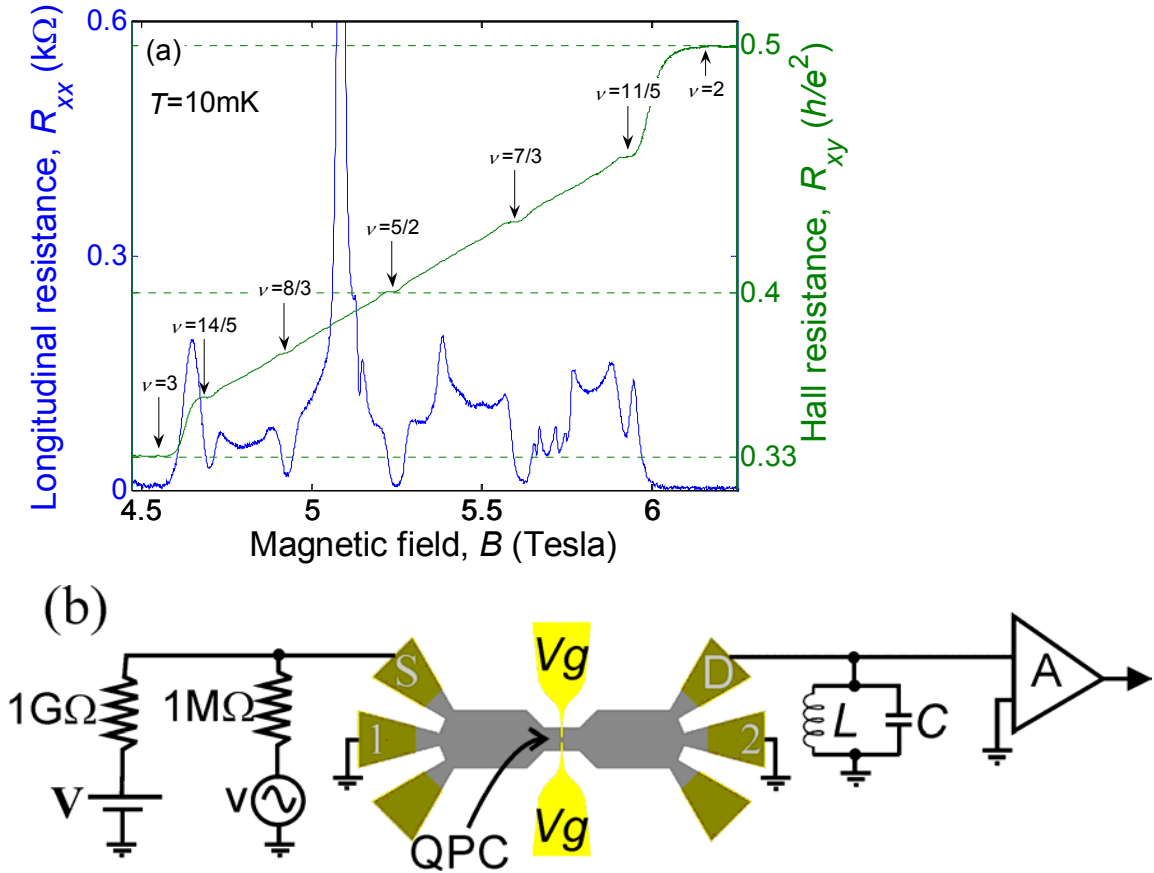

**Fig. 1.** Quantum Hall effect in the second Landau level. (a) Hall and longitudinal resistance measured on an ungated Hall bar, 1mm×2.5mm, with carrier density $3.15\times10^{11}\text{cm}^{-2}$ (determined at high magnetic field). The three main fractions, 8/3, 5/2 and 7/3, measured in a four terminal configuration on a Hall bar, are highlighted. (b) Schematics of the patterned sample. Carrier density in this sample was $3.27\times10^{11}\text{cm}^{-2}$. The two grounds, at 1 and 2, are 'cold grounds', cooling the electrons to ~10mK. DC current is driven to the sample through the source (S), provided by a DC voltage $V$ and a large resistor in series. The AC voltage $\upsilon$ is used to measure the conductance. Drain voltage (at D) is filtered with an *LC* resonant circuit, tuned to 910kHz, and amplified by a preamplifier cooled to 4.2K, adjacent to the sample. The quantum point contact, controlled by $V_g$, is tuned for the desirable transmission of the impinging current.

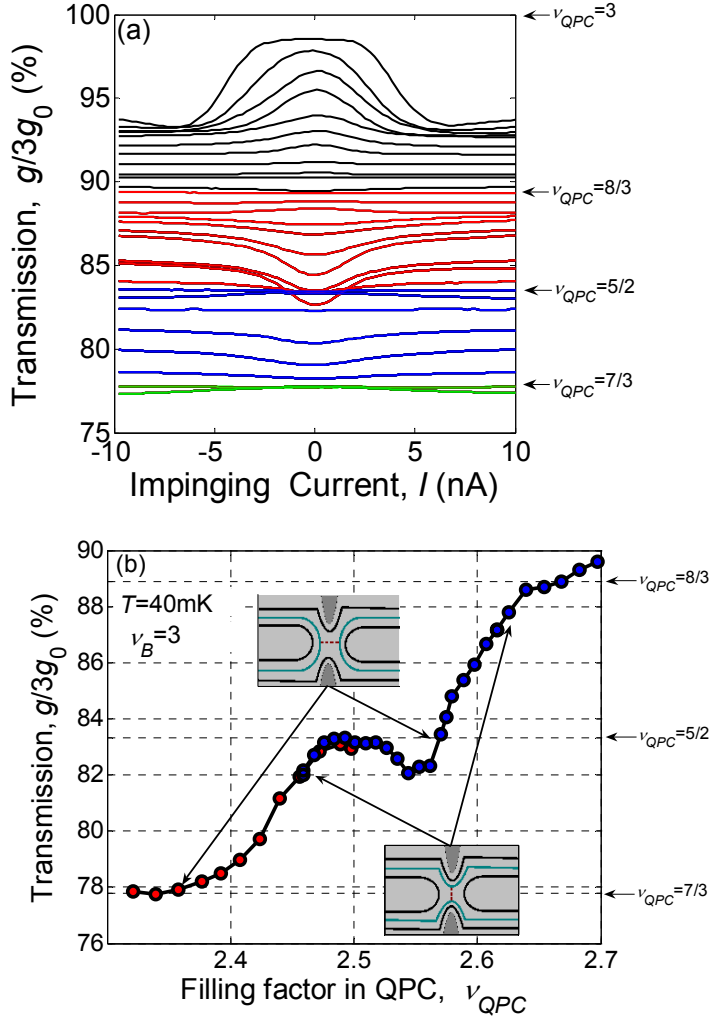

**Fig. 2.** The procedure used to identify lower lying states within the quantum point contact (QPC). (a) Differential conductance of the quantum point contact as a function of current at different filling factors in the QPC, which are adjusted by the gate voltage and the magnetic field (while staying on the $v_B=3$ plateau). Each fraction formed within the QPC is identified by a flat dependence of the conductance as function of current, surrounded by a 'mound-like' dependence for a slightly lower filling factor and a 'valley-like' slightly higher filling factor. The underlying states which were found using this method are 3: black for the partitioned state, 8/3 (red), 5/2 (blue), and 7/3 (green). The valley crossing near $v_{QPC}=5/2$ is due to the reentrant behavior to nearly an integer filling factor near that fraction. (b) Differential conductance of the quantum point contact as a function of the filling factors in the QPC. The two colors designate two different gate voltages, while at each gate voltage the filling factor was tuned with the magnetic field (on the $v_B=3$ plateau). The two insets describe schematically regions of weak backscattering (under a plateau) and regions of strong back scattering (just above the lower lying plateau).

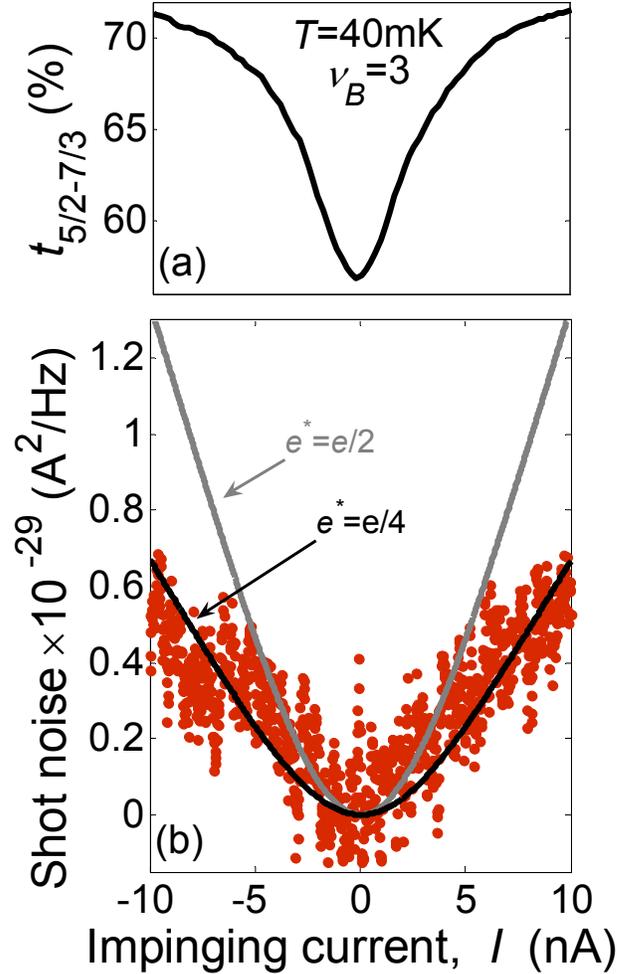

**Fig. 3.** (a) Conductance and (b) shot noise measurements of partitioned particles at 5/2 state. For a filling factor in the bulk $\nu_B=3$ and that in the quantum point contact tuned to weak backscattering of the 5/2 state, conductance and noise measured as function of the impinging current $I_{imp}=Vg_i$, with $g_i=3e^2/h$. The effective transmission of the partitioned channel was calculated assuming the lower state below the 5/2 is the 7/3 state (see Fig. 2). One thousand measurements points were taken during 40sec, as the impinging current changed from -10nA to 10nA. This measurement was repeated for a few hundred times, and then averaged. The amplification system, calibrated with a calibrated noise signal at 4.2K, had a voltage gain of 2,000. The predicted shot noise (Eq. 2) for charge $e^*=e/4$ (black line) and for $e^*=e/2$ (gray line) are plotted on the data. The data excludes the contribution of charge $e^*=e/2$ to the noise.

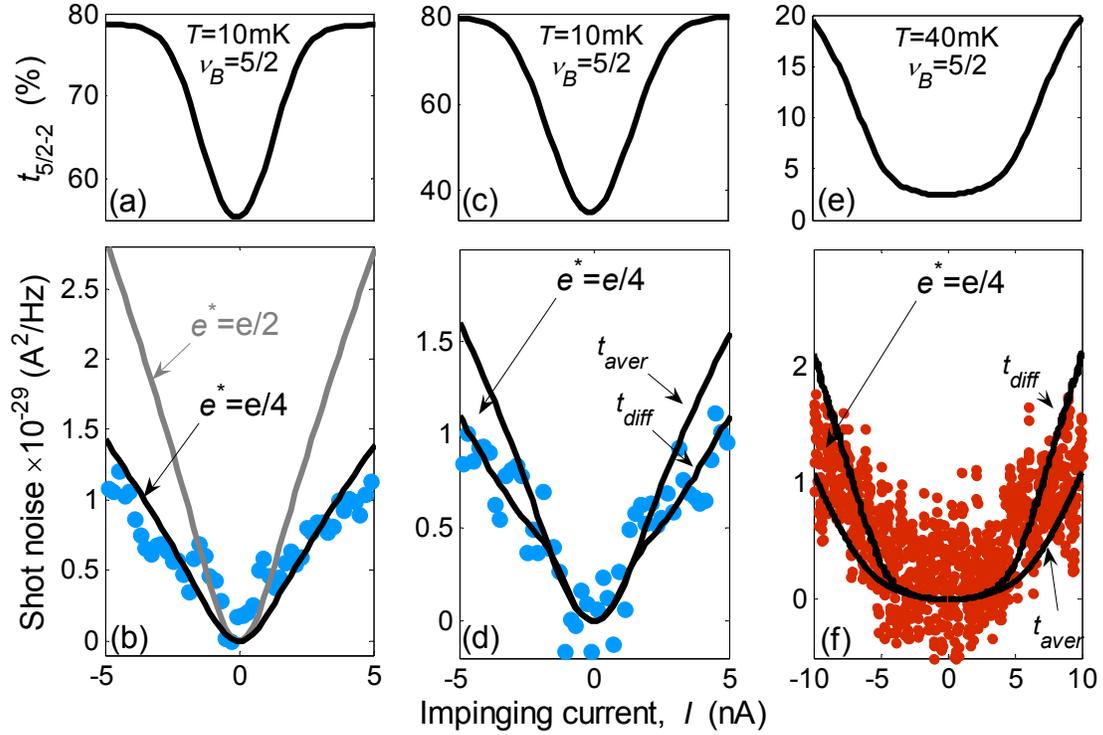

**Fig. 4.** Conductance and shot noise measurements of partitioned particles at 5/2 state, with a filling factor in the bulk $\nu_B$=5/2. Conductance and noise measured as function of the impinging current $I_{imp}=Vg_i$, with $g_i$=2.5$e^2/h$. The effective transmission of the partitioned channel was calculated assuming the lower state below the $\nu$=5/2 is the $\nu$=2 state, determined in a similar method to that described in Fig. 2. Measurements were done in a similar fashion to that described in Fig. 3, but some of the data is represented by less points, which were obtained after averaging. (a,b) Measurements at 10mK, at weak backscattering, where both models of the transmission coincide (see text). (c,d) Measurements at 10mK, transmission ~0.5, where the two models for the transmission provide the two limits of the expected noise. (e,f) Measurements at 40mK, strong backscattering, where the two models for the transmission provide the two limits of the expected noise. Surprisingly, the charge remains $e^*$=$e$/4, with no evident 'bunching' to $e$ when the QPC is nearly pinched.

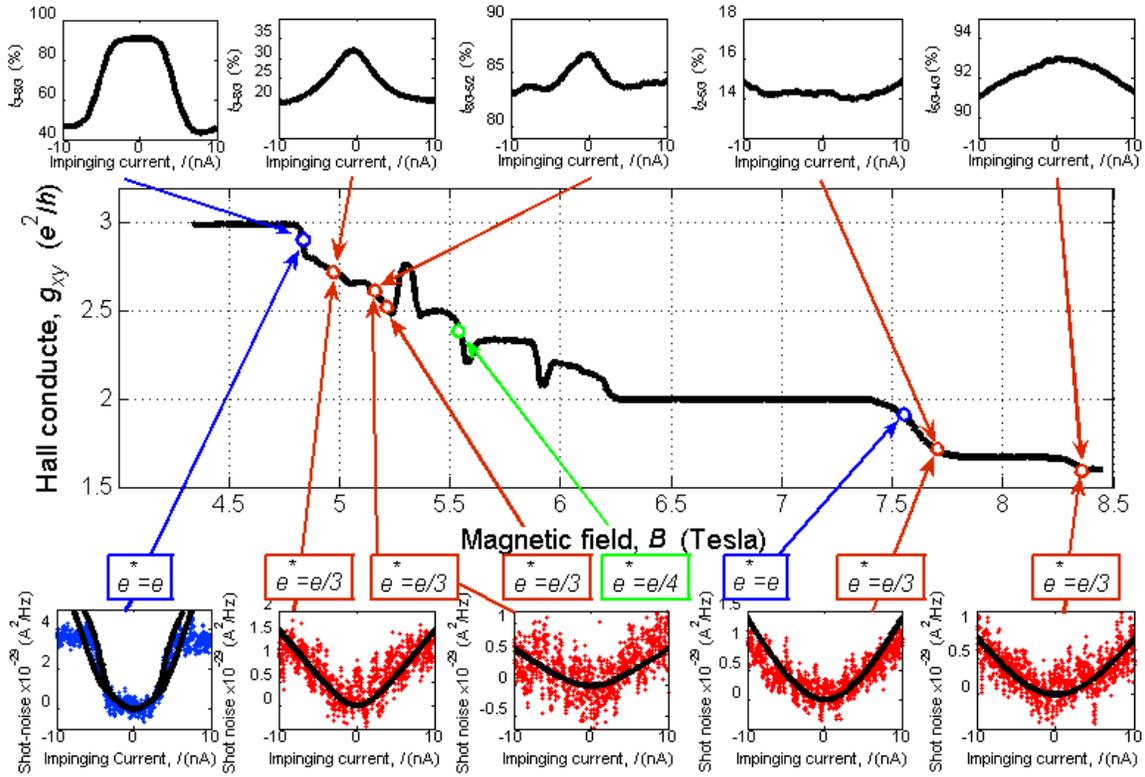

**Fig. 5.**
Conductance and noise in different filling factors in the quantum point contact. The measurement points are marked on the two terminal Hall conductance on the sample. Except for the 5/2 state, measurements in the range $\nu_{QPC}=3\ldots2$ were performed at $\nu_B=3$ and at $T=40$mK, and measurements in the range $\nu_{QPC}=2\ldots5/3$ were performed at $\nu_B=2$ and at $T=90$mK. Near integer fillings (2 & 3) the measured charge was $e$, while near (above and bellow) the fractional fillings (8/3 & 5/3) the measured charge was $e/3$.